\documentstyle[12pt,epsfig]{article}

\begin{document}

\title{Negative specific heat in a Lennard-Jones-like gas 
with long-range interactions}

\author{Ernesto P. Borges$^{a,b}$\thanks{ernesto@cbpf.br},
Constantino Tsallis$^a$
\\ $ $ \\
{\normalsize $^a$Centro Brasileiro de Pesquisas F\'\i sicas,} \\
{\normalsize R. Dr.  Xavier Sigaud 150, 22290-180 Rio de Janeiro--RJ, Brazil}
\\
{\normalsize $^b$Escola Polit\'ecnica, Universidade Federal da Bahia,} \\
{\normalsize R. Aristides Novis, 2, 40210-630 Salvador--BA, Brazil}
}

\date {}

\maketitle

\begin{abstract}
We study, through molecular dynamics, a conservative two-dimen\-sional
Lennard-Jones-like gas (with attractive potential $\propto r^{-\alpha}$).
We consider the effect of the range index $\alpha$ of interactions, number of
particles, total energy and particle density.
We detect negative specific heat when the interactions
become long-ranged ($0\le \alpha/d<1$).
\end{abstract}

\begin{quote}
\end{quote}

Non-Gaussian equilibrium in long-range Hamiltonian systems has recently
been reported \cite{Latora-Rapisarda-CT}. The authors studied a model of
classical planar spins with infinite range interaction, and showed 
that before going to the Boltzmann-Gibbs equilibrium state, the system 
reaches a metastable state with a non-Maxwellian distribution of
velocities. The duration of this metastable state appears to diverge 
with the number of particles, i.e., in the thermodynamical limit, 
the system indefinitely stays trapped in a non-Boltzmann-Gibbs state.

In this paper, we investigate similar anomalies with a different long-range
Hamiltonian model. We consider a two-dimensional gas confined in a box of
linear length $L$, defined by the Hamiltonian
${\mathcal H} = K + V/\tilde{N} + V_{walls}$,
where $K=\sum_{i=1}^{N} p_i^2/(2m)$,
the interacting potential is given by 
$V = \sum_{i<j}^N v(r_{ij})$ with \cite{Prausnitz}
$v(r_{ij}) = C_{\alpha}
\left[ \left( \sigma/r_{ij} \right)^{12}
- \left( \sigma/r_{ij} \right)^{\alpha} \right]$
$(0 \le \alpha <12)$
and
$C_{\alpha} = \epsilon \, (12^{12}/\alpha^\alpha)^{1/(12-\alpha)}/(12-\alpha)$.
$r_{ij}$ is the distance between the centers of particles $i$ and $j$,
$\sigma$ characterizes the diameter of a particle, $m$ is its mass 
and $\epsilon$ is the energy scale. We adopt appropriate units in which
$\sigma=\epsilon=m=1$. $C_\alpha$ is a factor that makes the depth of the
potential well the same $\forall \alpha$.
$V_{walls}$ is the repulsive potential of the walls, which assures that the
particles are confined within the box. We adopt soft walls, namely
$V_{walls} = \sum_{w=1}^{4} \sum_{i=1}^{N} (1/r_{iw})^{12}$,
where $w$ is the index for the walls, and $r_{iw}$ is the distance between
particle $i$ and wall $w$. 
The nonextensive scaling parameter 
$\tilde{N} \equiv 1+d \int_1^{N^{1/d}} dr \, r^{d-1-\alpha}$
(see \cite{CT:Springer}) is convenient to make the Hamiltonian \emph{formally}
extensive $\forall \alpha/d$ ($d$ is the spatial dimensionality of the system).
The model is a variation of the one introduced in \cite{Curilef}, and
the particular case $\alpha=6$ recovers the celebrated 
Lennard-Jones model \cite{Lennard-Jones} for dilute nonpolar gases. 
The variable involved in the range of interactions ($r_{ij}$) is also 
the dynamical variable, while in the classical spins model, 
the range of interactions is controlled by the (fixed) position of 
the particles in the lattice, and not by the angles, 
which are the dynamical variables.
We integrated the equations of motion using the forth order symplectic 
algorithm \cite{integrator} with time steps 
$\Delta t = 0.005$--$0.02$, which
yielded a maximum relative error in the total energy conservation
$\Delta E/E \sim 10^{-5}$. We adopted as initial conditions for positions
(a) all particles at a triangular lattice
(note that this geometry corresponds to 
the global minimum for a two-dimensional $L^2$ system), 
and (b) particles randomly distributed, and a water bag distribution for
initial velocities. 
We considered a long-range system with $\alpha/d=0.5$, and a short-range
$\alpha/d=3$, for comparison.

\begin{figure}[htb]
\begin{center}
\includegraphics[width=10cm,angle=-90,keepaspectratio]{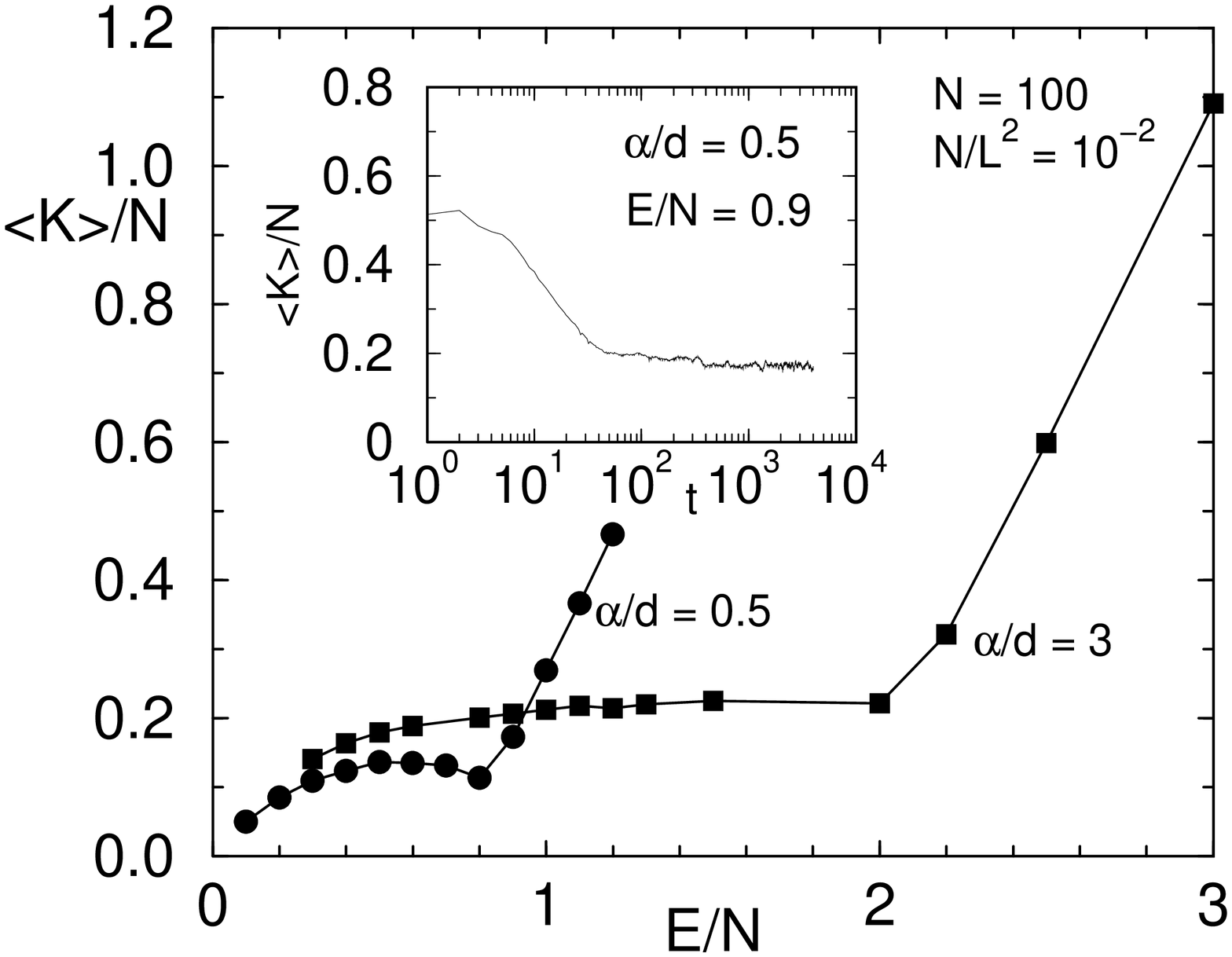}
\end{center}
\caption{\small Caloric curves for $N=100$ and $N/L^2=10^{-2}$:
$\alpha/d=3$ (squares), $\alpha/d=0.5$ (circles).
Solid lines are guides to the eyes. 
Inset: Typical time evolution of $\langle K \rangle /N$.}
\label{Fig_1}
\end{figure}

Fig.~\ref{Fig_1} shows caloric curves for short- and long-range systems
with $100$ particles. Each point of the caloric curve is obtained by a time
average after a transient of the order of $10^2$--$10^3$ time units 
(see Inset), and by an average of a few (up to 10) events with different 
initial conditions.
A negative specific heat region is exhibited for $\alpha/d=0.5$, while this 
effect does not appear in the short-range case.

\begin{figure}[htb]
\begin{center}
\includegraphics[width=10cm,angle=-90,keepaspectratio]{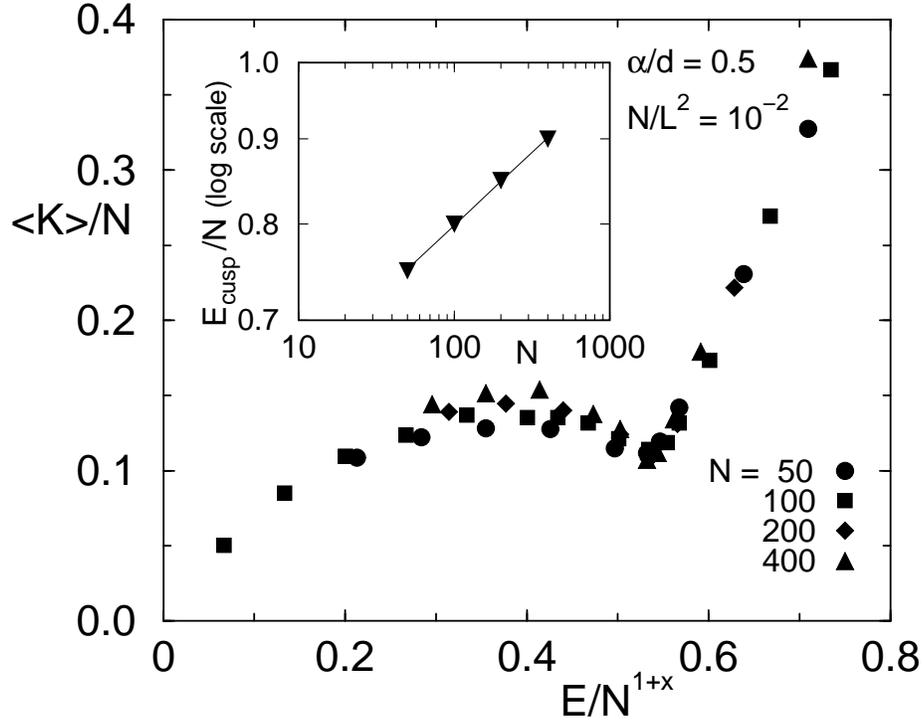}
\end{center}
\caption{\small Effect of the number of particles in the caloric curve, for the
long-range system ($\alpha/d=0.5$, $N/L^2=10^{-2}$).
$N=$ 50, 100, 200, 400. 
Inset: the energy per particle at the cusp point as a function
of $N$ ($\log$--$\log$ scale), slope $x=0.0876$.}
\label{Fig_2}
\end{figure}

Data collapse in the abscissa of the caloric curve (Fig.~\ref{Fig_2})
for the long-range case is achieved when the
total energy scales as $E/N^{1+x}$, with $x=0.0876$. Such small value of $x$
is not easily distinguishable (numerically speaking) from
a possible logarithmic correction hidden in the power-law scaling
introduced by $\tilde{N}$.
Such effect was not detected in the classical spins model
\cite{Latora-Rapisarda-CT}, and could be due to the boundary conditions.
Data collapse in the ordinate of the caloric curve may be roughly achieved
by scaling the maxima of the curves (for energies below the critical)
according to
$T_{max}(N) \sim T_{max}(\infty)-a/N^{y}$ ($T \equiv \langle K \rangle /N$)
with $y=0.038$, a value once again very small.

\begin{figure}[htb]
\begin{center}
\includegraphics[width=10cm,angle=-90,keepaspectratio]{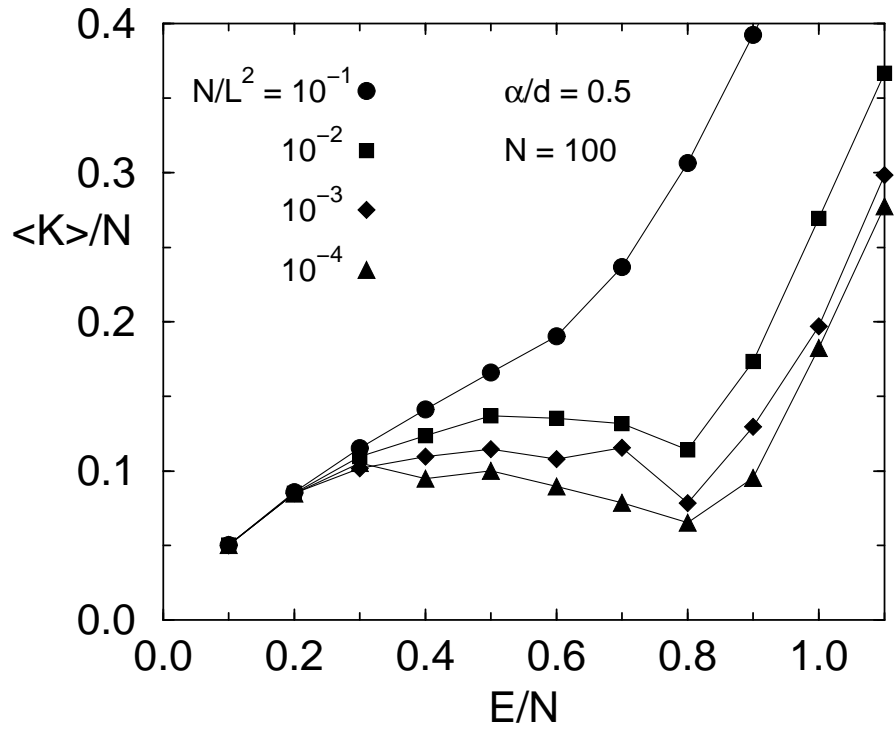}
\end{center}
\caption{\small Effect of the particle density in the caloric curve, 
for the long-range ($\alpha/d=0.5$) system with $N=100$.
$N/L^2=10^{-1}$, $10^{-2}$, $10^{-3}$, $10^{-4}$.}
\label{Fig_3}
\end{figure}

Fig.~\ref{Fig_3} displays the effect of the particle density in the 
caloric curve.
The negative specific heat region increases with decreasing density,
when long-range forces are present. Highly dense systems don't exhibit
such anomaly.

\begin{figure}[htb]
\begin{center}
\includegraphics[width=10cm,angle=-90,keepaspectratio]{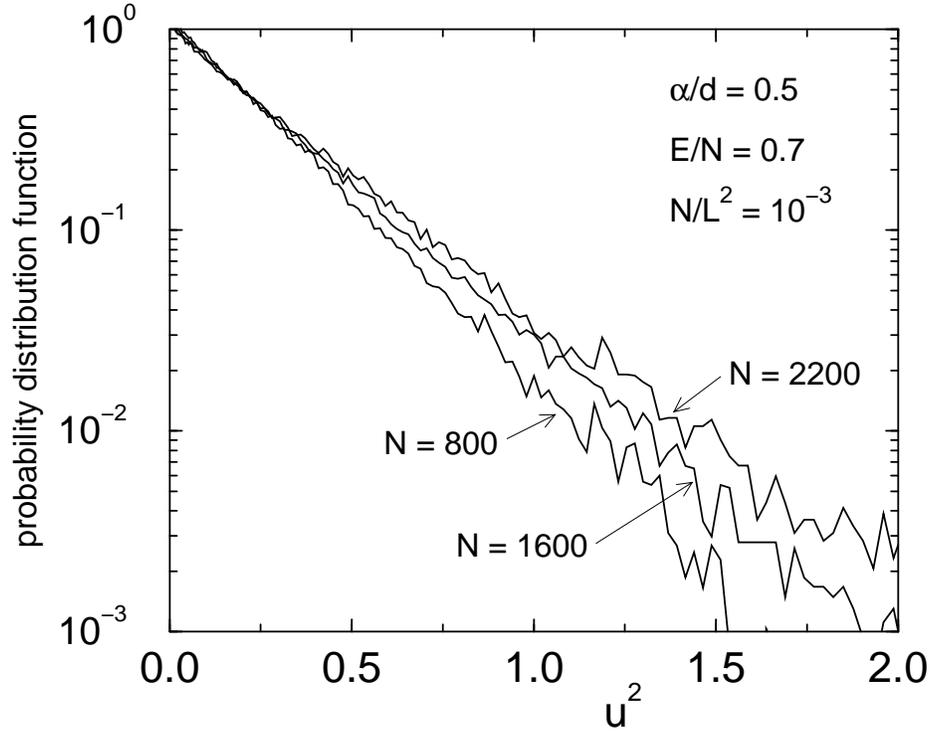}
\end{center}
\caption{\small Probability distribution function for the long-range case
($\alpha/d=0.5$), $E/N=0.7$ (below the critical energy per 
particle) and $N/L^2=10^{-3}$. $N=$ 800, 1600, 2200. 
We have taken, in the abscissa, the values of both $p_x$ and $p_y$.}
\label{Fig_4}
\end{figure}

Finally, Fig.~\ref{Fig_4} shows the probability distribution function (pdf) 
of velocities for the long-range ($\alpha/d=0.5$) system, with 
$N/L^2=10^{-3}$ and $E/N=0.7$
(a state in the negative specific heat region). As $N$ increases, 
the curves seem to departure from a Maxwellian,
exhibiting a slightly positive curvature.
A non-Maxwellian pdf was recently found in the classical spins model
\cite{Latora-Rapisarda-CT}, that might be understood within
nonextensive statistical mechanics 
\cite{CT:Springer,CT:1988,Curado_Tsallis,TMP}.
Such framework may also be the proper one to handle classical gases
with long-range interactions, but only further results would enable a
clear conclusion.

We warmly acknowledge the organizers of NEXT 2001 for the general support
and hospitality.
We also acknowledge CAPES, FAPERJ, CNPq, PRONEX/MCT 
(Brazilian agencies) for financial support.

\newpage

\newpage

\newpage

\newpage

\end{document}